\documentclass[]{aa} 
\usepackage[varg]{txfonts}
\usepackage[colorlinks=true,urlcolor=blue,citecolor=blue,linkcolor=blue]{hyperref}
\usepackage{graphicx}
\usepackage{natbib}
\usepackage{subcaption}
\usepackage{wasysym}
\usepackage{footnote}
\def\Halpha{\mbox{H$\alpha$}}
\def\kms{\mbox{km s$^{-1}$}}
\def\HeII{\ion{He}{II}}

\def\CaIIH{\ion{Ca}{II}~H}

\newcommand{\bea}{\begin{eqnarray}}
\newcommand{\eea}{\end{eqnarray}}
\newcommand{\grass}{}

\begin{document}

\title{Fan-shaped jets above the light bridge of a sunspot driven by reconnection}

\author{Carolina Robustini \inst{1} \and Jorrit Leenaarts \inst{1} \and Jaime de la Cruz Rodriguez \inst{1} \and Luc Rouppe van der Voort \inst{2}}

\institute{Institute for Solar Physics, Department of Astronomy,
  Stockholm University,
AlbaNova University Centre, SE-106 91 Stockholm Sweden \email{carolina.robustini@astro.su.se}
\and Institute of
  Theoretical Astrophysics, University of Oslo, P.O. Box 1029
  Blindern, N--0315 Oslo, Norway}

\date{Received; accepted}
\abstract

\abstract{We report on a fan-shaped set of high-speed jets above a strongly magnetized light
bridge (LB) of a sunspot observed in the \Halpha\ line. 
We study the origin, dynamics and thermal properties of the jets using
high-resolution imaging spectroscopy in \Halpha\ from the Swedish 1-m Solar Telescope and data from the Solar Dynamics Observatory and Hinode.
The \Halpha\ jets have lengths of 7--38~Mm, are impulsively
accelerated to a speed of $\sim 100$~\kms\ close to photospheric
footpoints in the LB, and exhibit a constant deceleration consistent
with solar effective gravity. 
They are predominantly launched from one edge of the light bridge,  and their footpoints appear bright in the \Halpha\ wings.
AIA data indicates elongated brightenings that are nearly co-spatial with the \Halpha\ jets. We interpret them as jets of at least transition region temperatures.
The magnetic field in the light bridge has a strength of $0.8-2~$kG and it is nearly horizontal. All jet properties are consistent with magnetic reconnection as the driver.}


\maketitle
\section{Introduction}

Complex sunspots can exhibit light bridges (LBs) crossing the
sunspot umbra. 
%
The chromosphere above LBs is rich in dynamic phenomena 
\citep[e.g.,][]{2003ApJ...589L.117B,2008SoPh..252...43L,2009ApJ...704L..29L}. 
Jets of cool ($<15$~kK) material are observed above certain LBs:
a detailed investigation of their dynamics is due to
\citet{1973SoPh...32..139R}
who extrapolated the full velocity vector of four jets, observed in \Halpha\, combining the plane-of-the-sky (POS) velocity of the jet fronts
and the magnetic field inclination, obtained from a potential field extrapolation 
\citep{Roy1973a}.
According to his analysis these jets can reach a maximum speed of 175 km~s$^{-1}$ and extend up to 50 Mm. They accelerate during an extended amount of time until they reach maximum velocity at a height between 7 and 50 Mm. Then they show a ascending phase where they decelerate stronger than the effective solar gravitational deceleration, and subsequently descending phase where the acceleration is smaller than the effective gravity. In order to explain such a behaviour \citep{Roy1973a} suggested a braking force, due to perturbation in the potential field, opposing the motion of the jets.

LB jets have been also studied more recently by other authors. 
\citet{asai}  
reported on dark recurrent \Halpha\ jets with apparent length of up to 23~Mm and visible in the extreme-ultraviolet channels of the Transition Region and Coronal Explorer (TRACE). These jets exhibit a fan-shape recognizable also in the observations of 
\citet{1973SoPh...32..139R}.
\citet{2007MNRAS.376.1291B} found mass ejections in a LB with a line-of-sight magnetic field that had opposite polarity  compared to the umbral field.
\citet{2009ApJ...696L..66S}
and
\citet{louis2014} 
observed shorter jets emanating from one edge of a LB with apparent length of
a few Mm using the wide-band \CaIIH\ filter on board the Hinode
spacecraft.
%
%
\citet{2015arXiv150502412B}
and
\citet{2015ApJ...811..138T}
reported on similar jets showing a parabolic motion in the POS. They both used 
observations from Hinode, the Interface Region
Imaging Spectrograph (IRIS) and the Atmospheric Imaging Assembly
\citep[AIA,][]{2012SoPh..275...17L}
on board the Solar Dynamic Observatory (SDO). 

In this paper we report on the dynamics and thermal structure of
recurrent jets above a LB with the same fan-shape, similar velocity
and apparent length as reported by 
\citet{1973SoPh...32..139R} and 
\citet{asai}.
We do so using imaging spectroscopy in \Halpha\ and imaging using
various AIA channels. Unlike
\citet{1973SoPh...32..139R},
we give the entire dynamic description independently of the inferred magnetic field vector.
Because of their fan shape and beauty in
\Halpha\, we dub them {\it peacock jets}. In lower resolution observations they 
are referred to as surges.

\section{Observations and data reduction}

 The jets were observed in AR11785 in its decaying phase on 2013 July 5 between 8:11 and
 9:38 UT with the CRisp Imaging SpectroPolarimeter
 \citep[CRISP,][]{2008ApJ...689L..69S} 
at the Swedish 1-m Solar Telescope
\citep[SST,][]{2003SPIE.4853..341S}.
Its heliocentric
coordinates were $(x,y)=(-485", -211")$ corresponding to an observing
angle of $32^\circ$ ($\mu = 0.85$). The measurement cadence was 8.8\,s
with 39 positions along the line profile between +95~\kms\ and
-95~\kms\ Doppler shift relative to line centre and a spacing of
5~\kms. The pixel scale was 0\farcs058. The CRISP data were reduced
following the pipeline described in
\citet{2015A&A...573A..40D}
which includes image restoration with Multi-Object Multi-Frame Blind Deconvolution
\citep[MOMFBD,][]{2005SoPh..228..191V}.
We also use co-aligned co-temporal data taken with the AIA and the
Helioseismic and Magnetic Imager
\cite[HMI,][]{2012SoPh..275..207S}
on board SDO. The cadence of the AIA and HMI was 12~s and 48~s, respectively. They have both been resampled to match the cadence and pixel
size of the SST data. 
The alignment was performed by first rotating the relevant SDO subfield to the same orientation as the SST data, and then performing a cross correlation between the SST \Halpha\ wide-band data and the HMI continuum data. The co-alignment is accurate to within an SDO pixel size of $0\farcs5$.
The SDO data show the jets were present from approximately 2013~July~4,~20:00~UT to~July~5,~22:00~UT.

In addition we use partially overlapping Hinode
\citep{2007SoPh..243....3K}
observations, consisting of a Solar Optical Telescope 
\citep[SOT,][]{2008SoPh..249..167T}
 wide-band Ca II H image taken at 09:23 UT and a SOT-spectropolarimeter fast raster in \ion{Fe}{1}\ 6302 taken between 8:14 and 8:46 UT.

We used the CRisp SPectral EXplorer
\citep[CRISPEX,][]{2012ApJ...750...22V}
software to analyse the data.

\section{Results}

\begin{figure*}
\includegraphics[width=17cm]{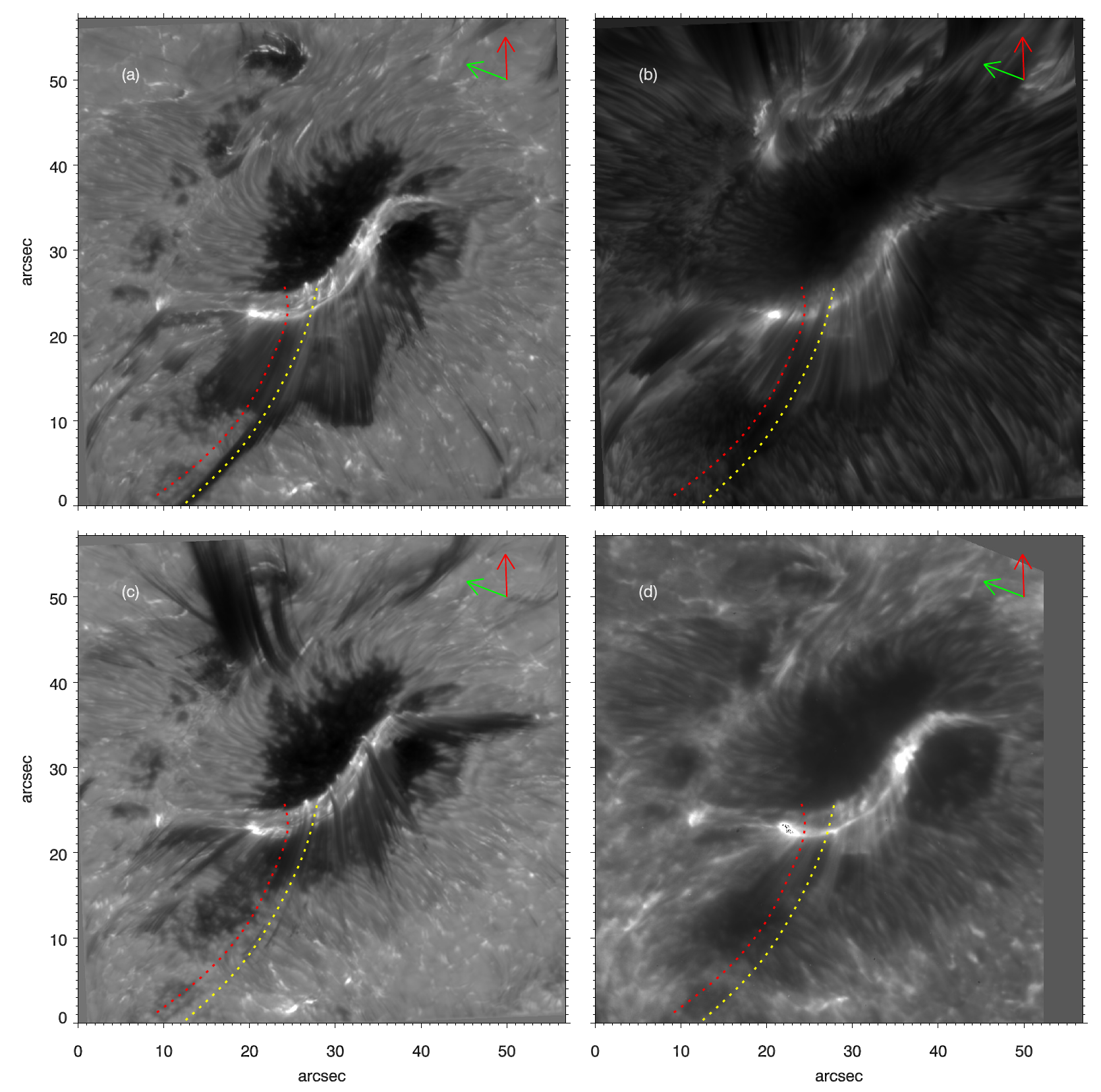}
\caption{Peacock jets observed in the H$\alpha$ line on 2013 June 5 at
  9:23~UT. (a) Blue wing image at $\Delta \lambda =$ -45~km~s$^{-1}$; (b) the line
  core; (c) red wing image at $\Delta \lambda =$ 45~km~s$^{-1}$; (d) co-temporal wide-band image in the SOT \CaIIH\ filter. The red and yellow dashed curves are the trajectories of the jets shown in Figure~\ref{fig:t_vs_s_slit5} and~\ref{fig:t_vs_s_slit5b} . The red and green arrows point towards the solar disk centre and the solar north pole, respectively. This figure is accompanied by an online animation that shows the temporal evolution of panel a-c.}
\label{fig:imm_bcr}
\end{figure*}

Figure~\ref{fig:imm_bcr} shows the peacock jets 
in the blue and red wing of \Halpha\ as well as in the line centre (a-c) and  \CaIIH\ (d). The
jets appear as slightly curved, dark absorption features in the blue
and red wing images respectively. They are less clear in the  \Halpha\  line core
image and in \CaIIH\ where they appear weakly in emission against the background of
the superpenumbra. The panels (a),(b) and (d) show a pattern of shorter and brighter jets launched together with the longer ones. The blueshifted and redshifted absorption features are not
co-spatial. The blue-wing
features appear predominantly rooted in a chain of finely structured
brightenings at the lower edge of the LB (Fig~\ref{fig:imm_bcr}a). 
In some cases the red-wing features extend to the upper edge of the LB (c). The \CaIIH\ jets have the same appearance as in 
\citet{2009ApJ...696L..66S}.

\begin{figure*}
\includegraphics[width=16cm]{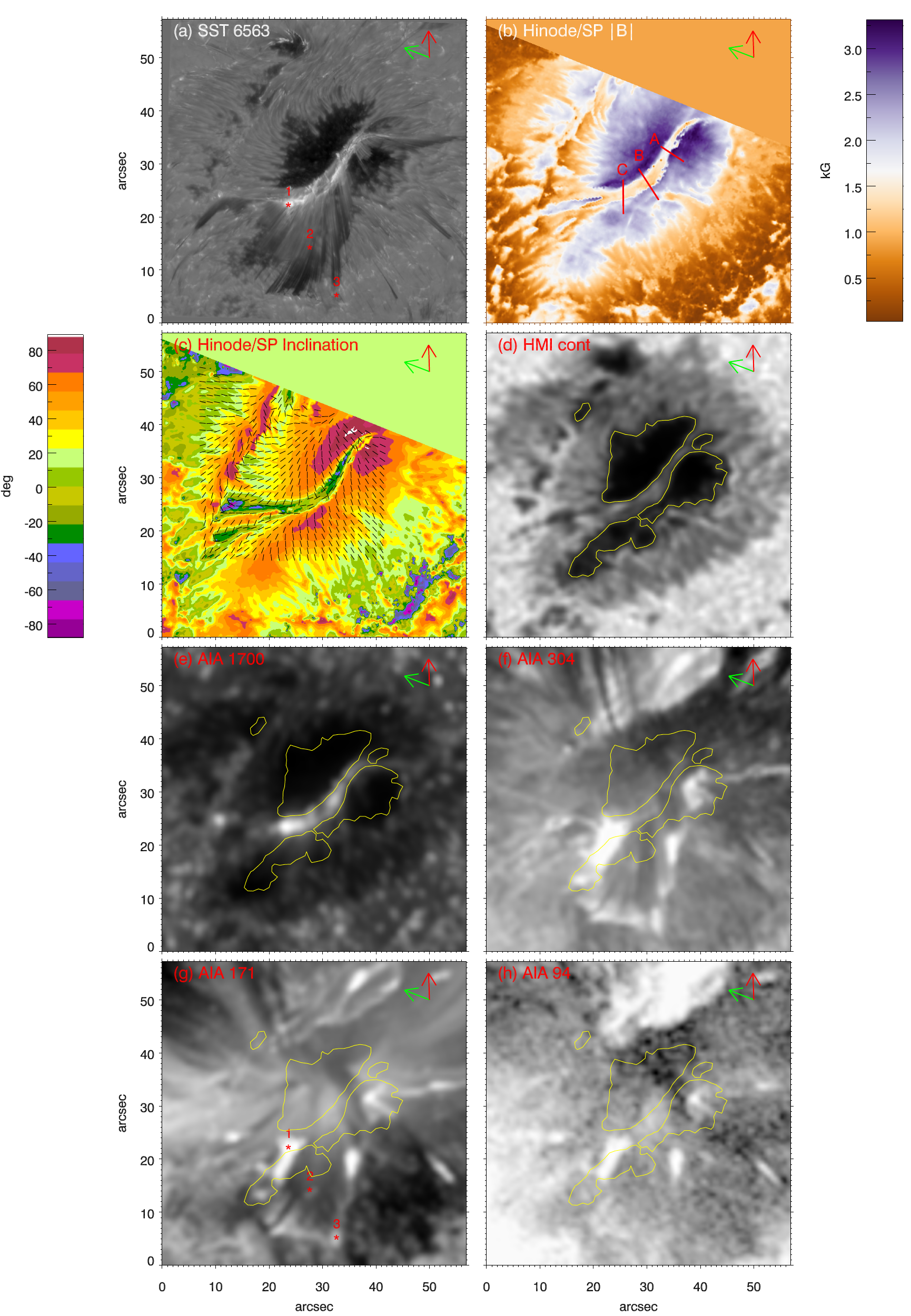}
\caption{(a) Peacock jets in the blue wing of H$\alpha$; (b) magnetic field strength $|\mathbf{B}|$. \grass
{The lines A, B and C refer to figure \ref{fig_b_profile}}; (c) magnetic field inclination  with respect to the solar vertical and the direction of the horizontal 
  component of the magnetic field (line segments). Panels (d)-(h) show
  images in the different AIA filters  co-temporal with the H$\alpha$ image. The yellow contour outlines the umbra. The numbered stars indicate points whose DEM is shown in panel (c) of
  Figure~\ref{fig:DEM}. This figure is accompanied by an online animation where the single Hinode SP scan (b) is replaced by the temporal evolution of the HMI magnetograms and the inclination map (c) by AIA~1600.}
\label{fig:imm_comeincrispex}
\end{figure*}

Figure~\ref{fig:imm_comeincrispex} compares the appearance of the
\Halpha\ jets with the magnetic field vector in the local reference system and the appearance in the AIA
channels. The field strength (b) and the local-frame inclination map (c) are obtained from an azimuth de-ambiguated inversion with a Milne-Eddington atmosphere done with MERLIN 
\citep{1987ApJ...322..473S}.
To solve the 180$^{\circ}$ ambiguity we applied the \textit{acute angle method} \citep{1985tphr.conf..313S}: we assumed a potential field and imposed that the LOS observed magnetic field and the transverse potential field components make an angle smaller than 90$^{\circ}$.

In the LB the field is rather horizontal and weaker ($0.8-2$~kG) compared to the more vertical umbral field which has a strength of $2-3.3$~kG.
The horizontal field component in the LB is aligned to the LB itself. A similar configuration is reported by \citet{2015ApJ...811..138T} but with a more weakly magnetized LB. The simulations of \citet{Toriumi2015b} suggest that the jets observed in \citet{2015ApJ...811..138T} can be driven by magnetic reconnection between the horizontal field carried into the LB and the vertical field of the surrounding umbra.
The LB field is weaker compared to the umbral field (2.2 - 3.2 kG). 

\begin{figure}
\includegraphics[width=\columnwidth]{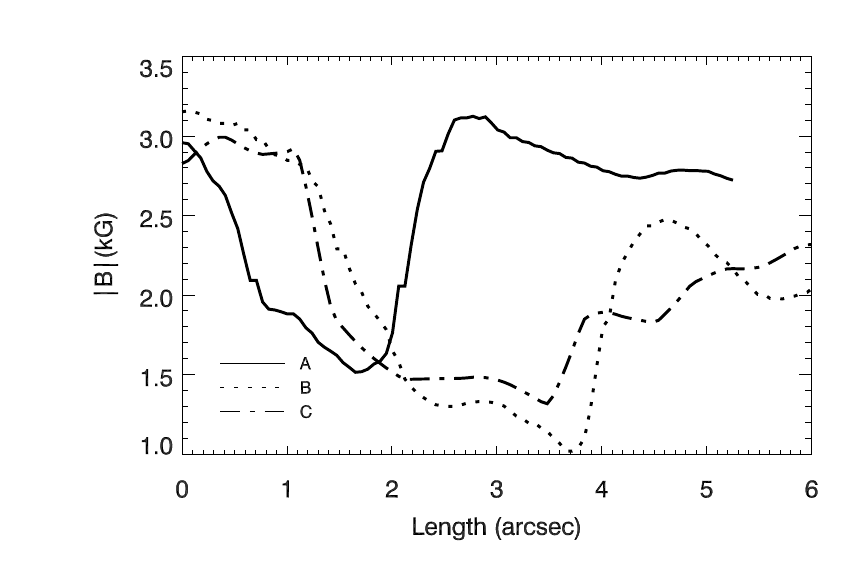}
\caption{\grass{Magnetic field strength profile plotted along the three lines A, B and C of Figure \ref{fig:imm_comeincrispex}-(b)}}
\label{fig_b_profile}
\end{figure}

\grass{Figure \ref{fig_b_profile} shows the profile of the magnetic field strength along three lines (A, B and C of Figure \ref{fig:imm_comeincrispex}-(b)) crossing the length of the LB at right angle.}

The bright jet footpoints are located where
the polarity of the $B_z$ field changes. The footpoints appear
bright in the 1700~\AA\ channel (and 1600~\AA) providing further
evidence that they are indeed located in the upper photosphere or
lower chromosphere
\citep{2005ApJ...625..556F}.
Some of the brightest footpoints (e.g. footpoint marked 1 in
\Halpha\ and 171~\AA) appear in all the AIA channels.
All channels display an elongated brightening co-aligned with the
direction of the \Halpha\ jet. All channels except
94~\AA\ continuously show a bright rim at the location of the top of
the \Halpha\ jets, and intermittently also in 94~\AA.

\begin{figure}[h]
\centering
\includegraphics[scale=0.9]{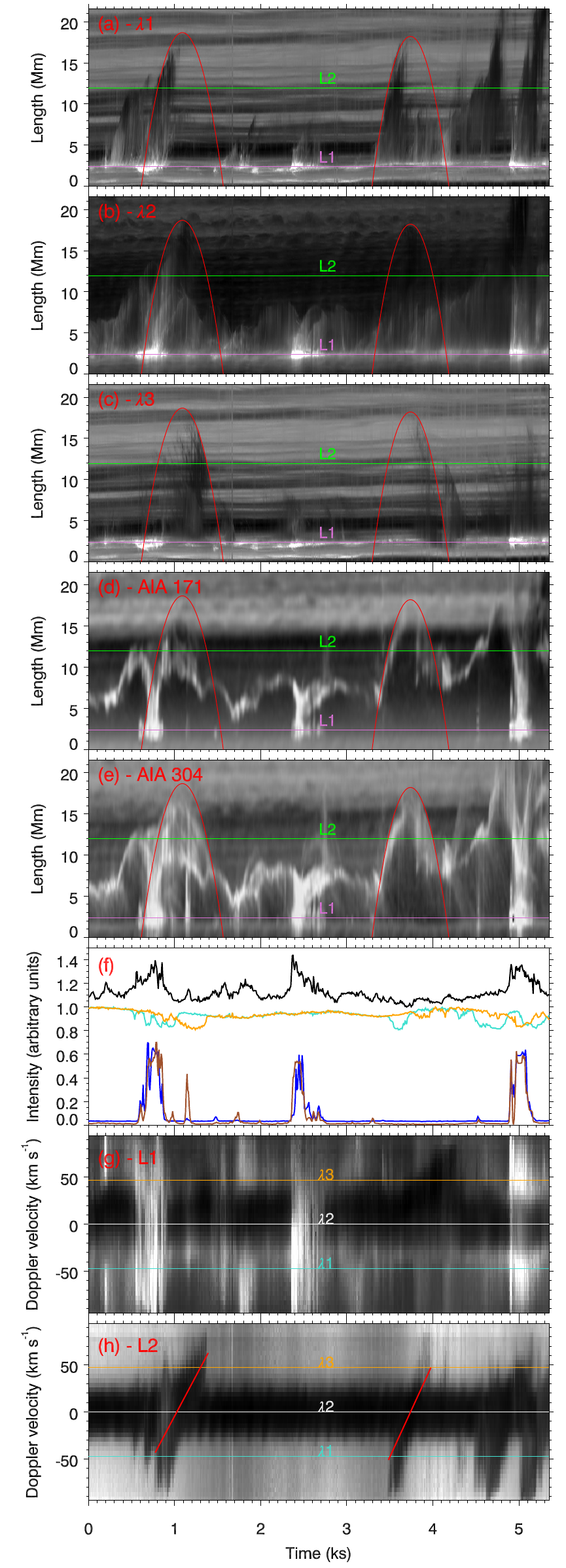}\caption{Panels a--c: time-slices ($s$-$t$ plots) along the red trajectory
  in Figure~\ref{fig:imm_bcr} for \Halpha\ -50~\kms\ (a),
  \Halpha\ 0~\kms\ (b), \Halpha +50~\kms \ (c), and two AIA channels
  (d--e). The red parabolas are fits to \Halpha\ jet fronts. 
  Panel f: normalized light-curves for \Halpha\  at $\lambda 1$ at location L2 (turquoise), 
  \Halpha\  at $\lambda3$ at L2 (orange), \Halpha\ at $\lambda 1$ at L1 (black), AIA 171 at L1 (blue), AIA 304 at L1 (brown). 
   Panels
  g--h: Time evolution of the \Halpha\ spectrum along the lines
  labelled L1 and L2 in panels a-e. The lines labelled
  $\lambda1$--$\lambda3$ are the wavelengths for which a time-slice is
  shown in panels a--c. 
  The red lines in panel h are fits to the Doppler shift of the absorbing feature associated with a jet.}
\label{fig:t_vs_s_slit5}
\end{figure}

\begin{figure}[h]
\centering
\includegraphics[scale=0.9]{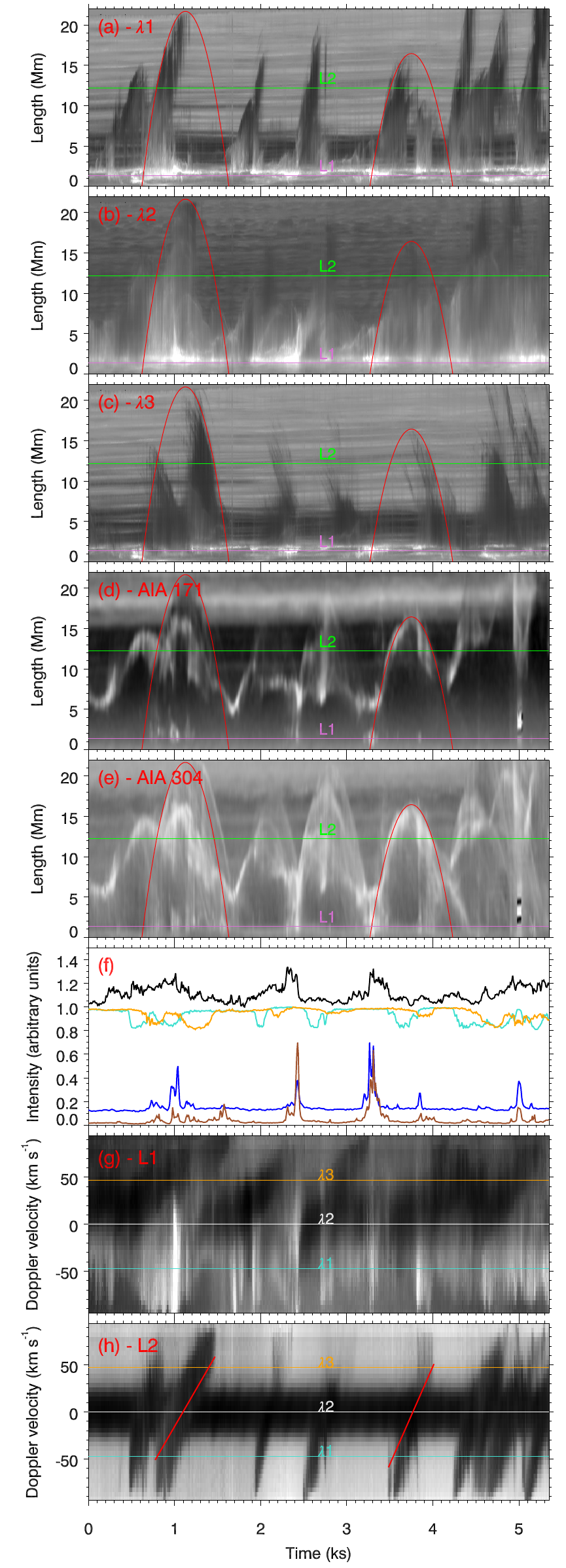}\caption{Same as in Figure \ref{fig:t_vs_s_slit5}, for the yellow trajectory
  in Figure~\ref{fig:imm_bcr}.}
\label{fig:t_vs_s_slit5b}
\end{figure} 
Figure~\ref{fig:t_vs_s_slit5} and \ref{fig:t_vs_s_slit5b} show space-time diagrams of the
intensity in \Halpha\ and AIA channels along the trajectories indicated
by the dashed lines in Figure~\ref{fig:imm_bcr}. 

Panels (a)-(c) show the evolution in the blue wing, line center and red wing of \Halpha. The
jets fronts appear as parabolic trajectories, with the upward part
most clearly seen in the blue wing, the tops in the line-center and
the downward part more visible in the red wing. The red curves
show a parabolic fit to the front of two jets. 

In Figure~\ref{fig:t_vs_s_slit5} the launch of the first jet coincides with a footpoint brightening at 0.7~ks, while for the second jet no intensity enhancement is visible on the LB. This is because the jets sometimes show sideways motion, and the trajectory misses the location of the brightening, but it is visible in the online animation of Figure~\ref{fig:imm_bcr}. At 2.6~ks a strong brightening is visible on the LB but no dark jet is associated to it, on the contrary we observe a shorter parabolic feature slightly in emission against the umbral background (see the animation at 08:50 UT).

The space-time diagram in Figure~\ref{fig:t_vs_s_slit5b}~(a) exhibits more prolonged brigthenings on the LB accompanied by three clear subsequent jets at 0.7~ks, 2.4~ks and 3.3~ks.
Most, but not all, jets emanate from brightenings at the lower edge of the light
bridge, marked with the position L1 in Figure~\ref{fig:t_vs_s_slit5} and~Figure~\ref{fig:t_vs_s_slit5b}. 

Panels (d)--(e) show the intensity in three AIA channels. The jet fronts
appear always bright in 171~\AA\ and 304~\AA. The three strongest \Halpha\ footpoint brightenings at 0.7~ks,
2.5~ks and 5.0~ks (Figure \ref{fig:t_vs_s_slit5}) coincide with bright jet-aligned streaks in all AIA
channels. The AIA channels in Figure~\ref{fig:t_vs_s_slit5b} do not show such dramatic events on the LB except the saturated pixels around 5~ks.
Panel f shows the light-curves for \Halpha\ , AIA 171~\AA\ and 304~\AA. The jet-aligned streaks in AIA (d)-(e) coincide with a sudden increase of intensity in the blue wing of \Halpha\ (black), in AIA 171~\AA\ (blue) and 304~\AA (brown). The peaks in \Halpha\ exhibit a slower decay than in the AIA channels. The jets appear as intensity drops when the jets move upward (turquoise, blue wing) and downward (orange, red wing) at position L2.

Finally, panels g and h show $\lambda-t$ plots of \Halpha\ along the
lines labelled L1 and L2 in panels a--c. The spectrum at the footpoints
shows many impulsive intermittent brightenings in the wings, that last approximately
fifty to several hundred seconds. Many, but not all, brightenings show
a corresponding subsequent parabolic jet profile in the space-time
diagrams. 
The stronger brightenings of Figure~\ref{fig:t_vs_s_slit5} coincide with the three intensity bumps in panel (g) of Figure~\ref{fig:t_vs_s_slit5}, which extend all along the \Halpha\ profile. On the contrary the bright footpoints of Figure~\ref{fig:t_vs_s_slit5b} have, as a counterpart, intensity bumps mostly in the blue wing.
Some jets associated with the other
brightenings follow slightly different trajectories and are thus
absent or partially visible. 
The three strongest brightenings have partially or completely filled-in
line-cores (see also Figure~\ref{fig:DEM}).

Panel h shows a rather standard \Halpha\ background profile. Strong
diagonal absorption features caused by the jets are superimposed with
maximum blueshifts and redshifts higher than $\pm 75$~\kms.  The
near-linear shift from blueshift to redshift indicates constant
deceleration of the jet material.

From Figures~\ref{fig:imm_bcr}--\ref{fig:t_vs_s_slit5} and the associated
movies the following qualitative picture arises: fast high-reaching
jets of plasma with temperatures lower than 15~kK, are continuously
launched during the full 1.5 hour duration of our observing sequence. Their footpoints are located at
the lower edge of a light bridge, where the field configuration (see panel c of Figure~\ref{fig:imm_comeincrispex}) suggests a magnetic shear between the LB horizontal field and the umbral field. They appear bright in \Halpha\ and AIA 1700~\AA\ and occasionally even in hotter AIA channels, and must
therefore be located in the photosphere or low chromosphere. The jets
are not launched at random times and locations. Instead, the footpoint
brightenings appear to ''ripple'' along the LB (see the online
movies). Once accelerated, the jet material follows a ballistic
trajectory and falls back to the solar surface. Some jets fall back to the upper edge of the LB, and Figure~\ref{fig:t_vs_s_slit5} shows evidence of almost co-spatial hot (AIA) and cool (\Halpha) jets. The presence of co-spatial cold and hot jets is consistent with simulations of flux emergence into pre-existing vertical field 
\citep[e.g.,][]{1995Natur.375...42Y}.
%
%
%

%
%
%
%

The jets develop a bright
front in the AIA channels that we tentatively interpret as the signature of
compressive density increase and heating as the jet rams into the
coronal material. However this brightening is shown also by the down-flowing 
material. This indicates that the physical process behind it is more
complex and requires further investigation.

We then proceeded to quantify the jet dynamics. We assumed that the
jets move along straight lines. Fig~\ref{fig:imm_bcr} and the
associated movie indicate that they do not. However, the deviations
from a straight line are not large, so the error we make is likely to be small
compared to the accuracy with which we can measure the velocity and
deceleration of the jets.

We traced the POS of all jets that show a clearly defined ascending and descending trajectory (40 jets in total) in
\Halpha\ and extract the $\lambda$-$s$-$t$ data along the path. Then,
based on $s$-$t$ cuts we fitted the POS motion of the jet front with a
parabola (examples are shown in panels a--c of
Figure~\ref{fig:t_vs_s_slit5} and Figure~\ref{fig:t_vs_s_slit5b}). We then extracted $\lambda$-$t$ slices
for several positions along the parabola and measure the line-of-sight
(LOS) velocity of the jet front from the maximum Dopplershift of the
absorption feature (as in panel~h of Figure~\ref{fig:t_vs_s_slit5} and~\ref{fig:t_vs_s_slit5b}),
while we obtained the LOS deceleration from the slope of the absorption
feature in the $\lambda-t$ slice. Using those measurements we fitted a
linear function to the velocity as function of time of the jet front
along the LOS. Based on the dynamics of the jet fronts along the
LOS and in the POS, we determined their distance from the
photospheric footpoint $s$ as function of time, the initial velocity
$v_0$ and deceleration $a$:
\bea
s(t) &=& v_0 (t-t_0) -  \frac{a}{2} (t-t_0)^2 \\
v_0 &=& (v_\mathrm{LOS}^2(t_0) +v_\mathrm{POS}^2(t_0))^{1/2} \\
a &=& (a_\mathrm{LOS}^2 +a_\mathrm{POS}^2)^{1/2},
\eea
where $t_0$ is the time the fitted parabola along the POS motion
crosses the bright edge of the light bridge.  The initial velocity is
thus an estimate of the jet velocity directly after the impulsive
acceleration.

From the position of the jets on the solar disk and the POS and LOS
velocities we computed the angle $\theta$ of the jet with the solar
vertical assuming motion along a straight line using the procedure
outlined in
 \citet{1993ApJ...403..780T}, 
which allows us to compare the deceleration of the jet to the
effective gravity $g \cos{\theta} $, with $g$ the solar gravitational
acceleration.

 \begin{figure}
\includegraphics[width=\columnwidth]{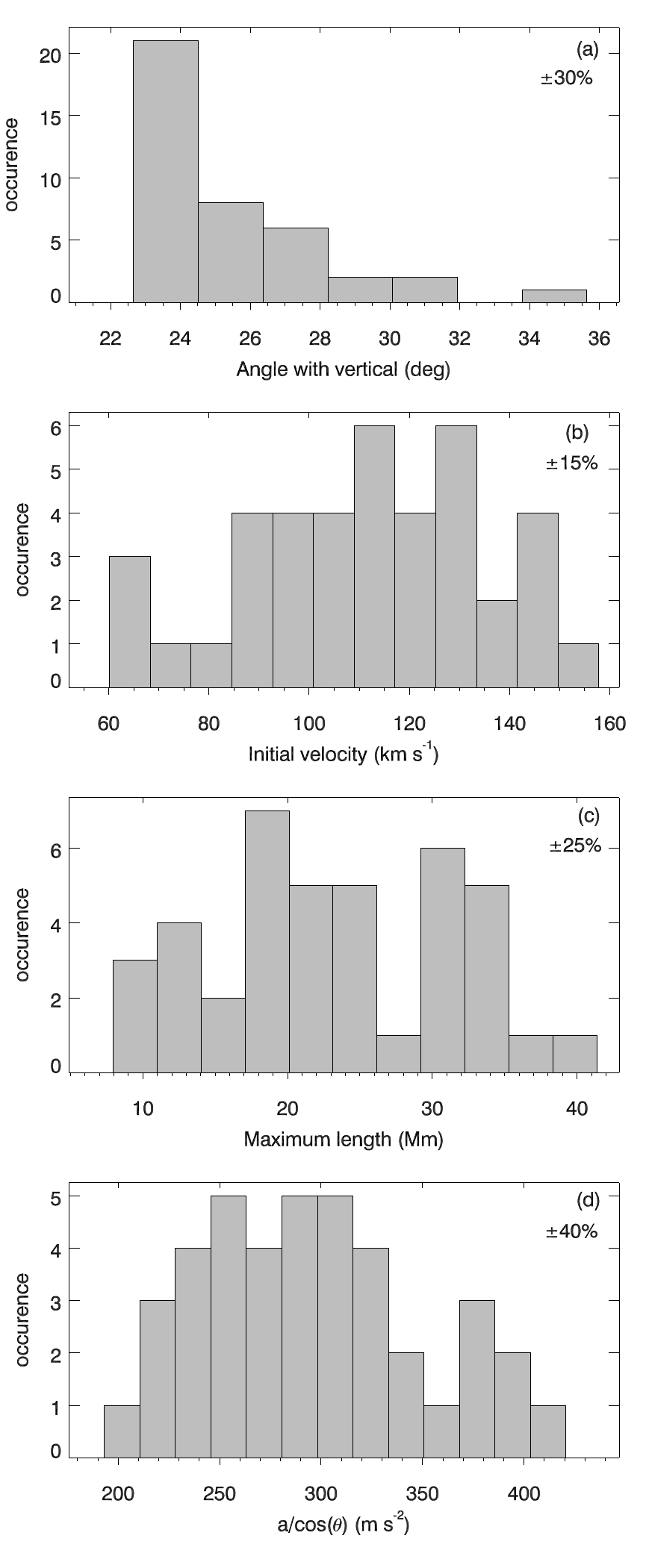}
\caption{Histogram of jet properties. The sample size is $N=40$. Relative
  errors are given in the upper right corner of the panels.  (a) angle $\theta$ with the solar
  vertical; (b) inferred initial jet velocity; (c) maximum jet extent;
  (d) $a/\cos{\theta}$, the post-launch deceleration divided by the cosine of the angle with the solar
  vertical.}
\label{fig:histograms}
\end{figure}

In Figure~\ref{fig:histograms} we show histograms of the jet angle
with the vertical, initial jet velocity, maximum jet length and $a/\cos{\theta}$. The latter quantity would be equal to the solar gravity  if gravity is the only force acting on the jet.
The jets make an angle of
~25$^{\circ}$ with the LOS. Owing to the large thermal width of the
\Halpha\ line, and the superposition of multiple jets along the LOS we
predict typical errors of 10~\kms in the determination of the Doppler
shift of the absorption features in the spectra, and 15\% relative
error in the LOS deceleration. The POS velocity and deceleration can
be determined with 5\% error. This leads to errors in the quantities
shown in Figure~\ref{fig:histograms} ranging from 15\% in $v_0$ to
40\% in  $a/\cos{\theta}$. The initial velocities are
typically ten times the local sound speed or larger. The lengths are
consistent with those reported by
\citet{1973SoPh...32..139R}
and
\citet{asai}.
The decelerations are consistent with solar effective gravity without
significant other forces.
 
 \begin{figure}
\includegraphics[width=\columnwidth]{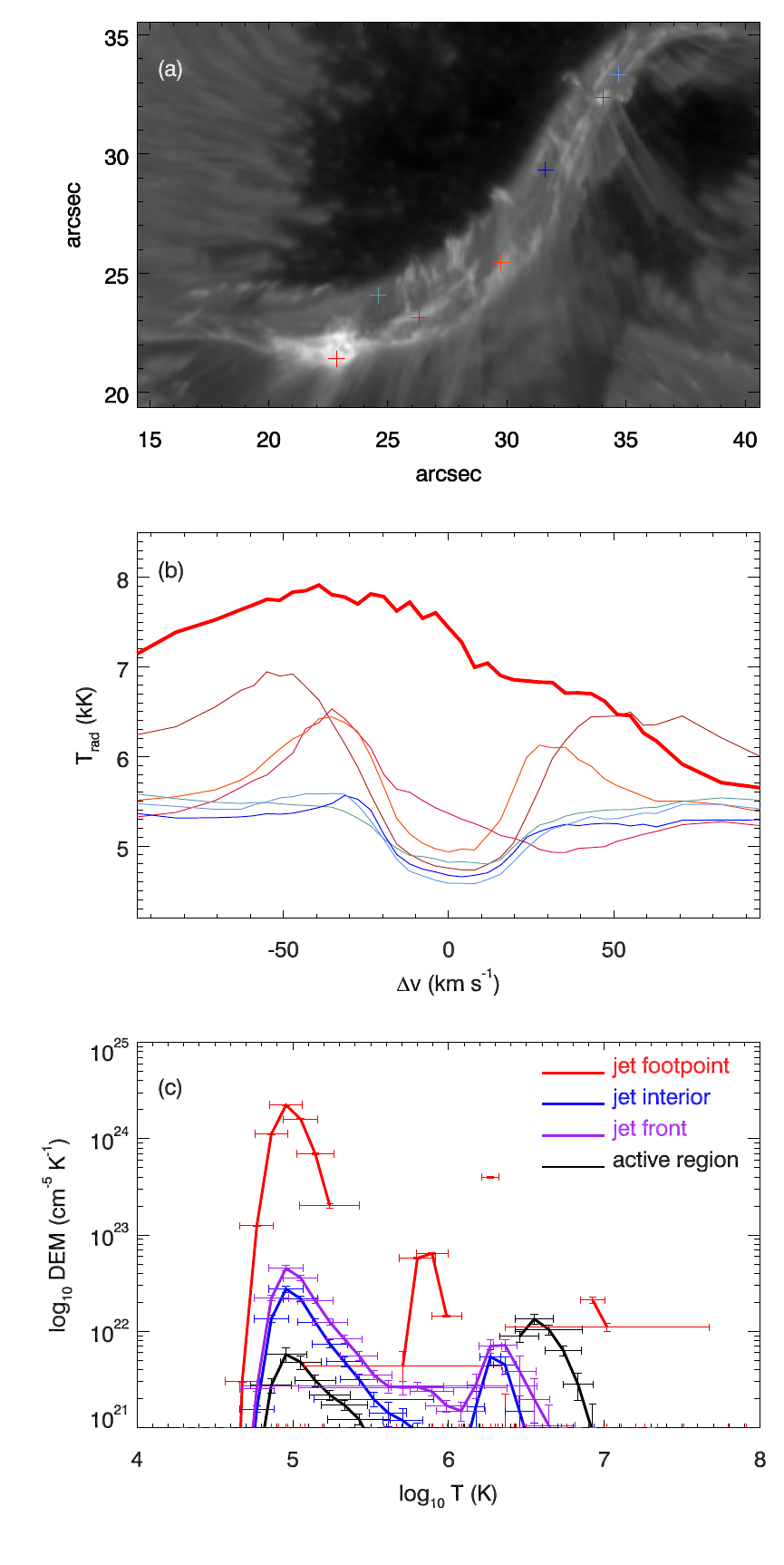}
\caption{Top: \Halpha\ image of the LB at $\Delta v=-50$~\kms, with
  locations of the line profiles in the middle panel indicated with
  plus-symbols. Middle: typical \Halpha\ profiles in the dark part of
  the light bridge (shades of blue), and typical profiles in the
  bright jet footpoints (shades of red). The thick red curve is the
  profile of the pixel labelled 1 in
  Figure~\ref{fig:imm_comeincrispex}. Bottom: DEM distribution for the
  pixels labelled 1--3 in Figure~\ref{fig:imm_comeincrispex} and an
  average active region area.}
\label{fig:DEM}
\end{figure}

 Finally, we investigated the thermal properties of the acceleration sites. 
 The jet footpoints are not bright in the HMI continuum (see
 Figure~\ref{fig:imm_comeincrispex}), suggesting no or weak heating in
 the deep photosphere, but appear bright in AIA 1700~\AA. We computed
 the brightness temperature $T_\mathrm{B}$ of the \Halpha\ data using
 calibration of a quiet region in the FOV with an atlas profile
\citep{1984SoPh...90..205N}
and correcting for solar limb darkening
\citep{1994SoPh..153...91N}.

Figure~\ref{fig:DEM} shows example profiles in the LB. Dark places
typically show brightness temperatures of 5.5~kK in the wings, and a
normal unshifted line core. The line profiles in the jet footpoints
are very different. They typically have bright wings ($T_\mathrm{B}
\sim 5.5$~kK---$7$~kK) and a wide core, somewhat reminiscent of
''moustache profiles" in Ellerman bombs
\citep{1917ApJ....46..298E}. 
The line core absorption is caused by overlying jet material
(Figure~\ref{fig:imm_comeincrispex}). The strongest brightenings have a
$T_\mathrm{B}$ up to 8~kK, and show weak redshifted line-cores and
maximum emission on the blue side of nominal line center. Assuming
that the brightness temperature is a fraction of the gas
temperature, which is reasonable given a source function with a
scattering and a thermal component, then the foot-point material
visible in \Halpha\ is hotter than 8~kK. The strongest
\Halpha\ brightenings are also visible in the AIA channels, typically
as the footpoint of an elongated feature (such as the point marked
''1'' in Figure~\ref{fig:imm_comeincrispex}). We therefore used the
AIA channels to compute the differential emission measure (DEM) using
the method described in
 \citet[][\footnote{code available at \url{http://www.astro.gla.ac.uk/~iain/demreg/}}]{2012A&A...539A.146H}.
The limited number of AIA channels and their wide temperature response
limits the accuracy of the DEM analysis. An additional complication is that
the \HeII~304~\AA\ line dominating the AIA 304 channel does not form
under coronal equilibrium conditions
 \citep[see for example][]{2014ApJ...784...30G},
limiting the reliability at temperatures below 100~kK. Nevertheless
 we show the results for the bright point labelled ''1'' in
 Figure~\ref{fig:imm_comeincrispex} in the bottom panel of
 Figure~\ref{fig:DEM} and compare it with a typical DEM distribution for
 the inside of a jet ("2"), the jet front ("3") and the average over an
 area in the lower left corner of the SST FOV. The footpoint DEM is
 two orders of magnitude higher at a temperature of 100~kK. This is
 suggestive of transition region temperatures down at the acceleration
 site in extreme cases. The AIA brightening at the jet front has a
 higher than typical DEM, supporting the idea of compression of
 coronal material.

\section{Discussion \& Conclusions}

We have investigated recurrent jets (that we dubbed {\it peacock jets}) of
cool ($<15$~kK) material that are launched along the light bridge of a
sunspot in AR11785. Using imaging spectroscopy we measured both the
LOS and POS components of the velocity and deceleration of the
jets. The latter are impulsively accelerated in the upper photosphere and/or
chromosphere above the sunspot to speeds up to 145~\kms, decelerate at
solar effective gravity and reach lengths of up to 40~Mm. The jets
have an angle of about 25$^\circ$ with the solar vertical. 
The ballistic motion of the peacock jets is not in agreement with the results of
\citet{1973SoPh...32..139R},
who observed extended acceleration followed by a braking stronger than gravity and eventually a fall towards the solar surface with an acceleration smaller than gravity.
The reason for this discrepancy can reside in a different approach: (1) we do not need to assume a potential field to retrieve the LOS velocity component; (2) we have a sample ten times larger, (3) our data have a higher resolution and higher cadence; (4) our deceleration uncertainty is obtained from a proper error analysis. We believe that the results obtained by 
\citet{1973SoPh...32..139R} 
are affected by small number statistics and an overly optimistic error estimate, and note that the curves in his Figure~2 could reasonably well be fitted with a parabola.


The accuracy of our derived velocity and deceleration is mainly set by
the accuracy with which the Doppler shift of the jet material can be determined. The POS component of the motion can be derived with
smaller errors. It would therefore be of value to observe peacock jets
close to the limb, where the jet axes are more aligned with the POS to
derive more accurate values of the deceleration and so check whether
gas pressure gradients or other forces significantly lower the deceleration as it does in
shock-driven dynamic fibrils in plage
\citep{2007ApJ...655..624D}
and the chromosphere of sunspots
\citep{2013ApJ...776...56R}.

The jet fronts appear bright in the AIA 304, 211, 193, 171 and 131
channels and intermittently even in the 94~\AA\ channel. Supported by
our DEM modelling we tentatively interpret this as the effect of shock compression of
transition region and coronal material as the cool jet slams into
the corona. This is in contrast with
\citet{asai} 
who interpret the bright rim seen in 171~\AA\ around the \Halpha\ jets
as evidence of a magnetic loop. We find this scenario unlikely as it
would require the \Halpha\ jets to move at right angles with the
magnetic field.

The most interesting aspect of the jets is their acceleration
mechanism. Jets are launched with speeds in the range 70--145~\kms,
which is nearly 10 times the sound speed at a temperature of
10~kK. These high speeds are hard to reconcile with magneto-acoustic
shock driving. Shock driven events such as dynamic fibrils
exhibit much lower velocities
\citep[up to 35~\kms, see][]{2006ApJ...647L..73H,2007ApJ...655..624D,2013ApJ...776...56R}.
On contrast,  speeds of order 100 km~s$^{-1}$ are found in small-scale penumbral jets \citep{2007Sci...318.1594K} which are driven by reconnection between the horizontal dark penumbral filament field and the more vertical bright penumbral field.
The peacock jets are mainly launched from the edge of the
light bridge, where the magnetic field changes rapidly from vertical to horizontal. Magnetic reconnection thus appears to be the
only viable candidate mechanism. 
The following facts also support the reconnection scenario:
the acceleration site heating is impulsive as evidenced by the sudden increase in \Halpha\ brightness at the footpoints;
the jets are predominantly launched from one side of the LB only. This is consistent with the asymmetry induced by the slightly inclined umbral field;
we observe jet material flowing down to the other edge of the LB, in agreement with models where cool jet material can slide down along post-reconnection field lines
\citep{1995Natur.375...42Y, yokoyama1996, moreno2008}.

We do not believe that peacock jets are caused by submergence of
umbral field by LB convection, a scenario proposed by
\citet{2014A&A...568A..60L} for granular LBs,
as this process would lead to jet acceleration at both LB edges, which
is not observed.
We are convinced that the data we analysed show
essentially the chromospheric and coronal aspects of the phenomenon
observed by
\citet{2009ApJ...696L..66S}
and
\citet{louis2014}.
These authors observe short jets above sunspot light bridges using
Hinode SOT \CaIIH\  and study the corresponding photospheric magnetic
field using observation with Hinode SP.
\citet{2009ApJ...696L..66S}
speculate that the LB could harbour a twisted
horizontal flux tube embedded below the canopy
formed by the strong umbral field. But, given the constraints of the resolution and the atmospheric model, the inclination map of Figure~\ref{fig:imm_comeincrispex}-(c) do not show any evidence of a twisted flux tube.

The observed LB and its field configuration are roughly reproduced in the simulations of \citet{Toriumi2015b}. In their simulation it is the result of a convective upflow carrying a horizontal field to the surface. The horizontal field is aligned along the LB and at the boundaries the shear with the pre-existing vertical field can give rise to magnetic reconnection. Our peacock jets observations are consistent with this scenario.

\grass{We note that the LB exhibits a filamentary structure (see Figure \ref{fig:imm_bcr}) somewhat similar to penumbral filaments and penumbral intrusions. The chromosphere above the latter two can also show jets
\citep{2007Sci...318.1594K,2015arXiv150902123B}.
 }


The appearance of the acceleration sites in \Halpha\ and the SDO
channels is remarkably similar to that of Ellerman bombs
\citep{2013ApJ...774...32V,2013JPhCS.440a2007R},
so lessons learned from Ellerman bombs might be applicable to the
acceleration of peacock jets. Coordinated SST and IRIS spectroscopy indicate that Ellerman bombs are sites where the photospheric gas underneath the chromospheric canopy is heated sufficiently to momentarily reach stages of ionization normally assigned to the transition region and the corona 
\citep{2015ApJ...812...11V}.
We speculate that equally vigorous reconnection might happen at the peacock jet footpoints, leaving a similar imprint in the line profiles. The acceleration sites share similarities also with the hot explosions described in 
\citet{peter2014hot}:
magnetic reconnection taking place in the upper photosphere heats the surrounding plasma to temperatures normally associated with the transition region (but see  
\citet{2015ApJ...808..116J})
with cold plasma on top. Ellerman bombs and the events reported in \citet{peter2014hot} are not associated with clear high-reaching upward moving jets, in contrast to peacock jets. Most likely the vertical magnetic field above the umbra allows for efficient jet formation, while the horizontal magnetic canopy above Ellerman bombs and hot pockets inhibits jet formation.

\begin{acknowledgements}
The Swedish 1-m Solar Telescope is operated by the Institute for Solar
Physics of Stockholm University in the Spanish Observatorio del Roque
de los Muchachos of the Instituto de Astrof\'{\i}sica de Canarias.
Hinode is a Japanese mission developed by ISAS/JAXA, with the NAOJ as domestic partner and NASA and STFC (UK) as international partners. It is operated in cooperation with ESA and NSC (Norway). 
This research has benefited from discussions at the International
Space Science Institute (ISSI). We thank Jayant Joshi for his valuable suggestions. Jaime de la Cruz Rodr{\'i}guez acknowledges support from the Swedish Research Council and the Swedish National Space Board. 
\end{acknowledgements}



\end{document}